
%
%
%

\font\bbigten=cmr10 scaled\magstep2

\def\bra#1{\displaystyle\langle\, #1\,|}
\def\ket#1{\displaystyle|\, #1\,\rangle}

%
\def\qad@rato#1#2{{\vcenter{\vbox{
    \hrule height#2pt
    \hbox{\vrule width#2pt height#1pt \kern#1pt \vrule width#2pt}
    \hrule height#2pt} }}}
\def\qadratello{\mathchoice
 \qad@rato{5}{.5}\qad@rato{5}{.5}
 \qad@rato{3.5}{.35}\qad@rato{2.5}{.25} }
%
%
\magnification=1200
\baselineskip=8truemm plus2truemm minus1truemm
\vsize=22.1truecm
\hsize=15truecm
\hoffset=0truecm
\null
\bigskip
\centerline{\bbigten The Accelerated Observer and Quantum Effects}
\medskip
\centerline{Roberto Casadio{\footnote{$^\dagger$}{e-mail:
Casadio@bologna.infn.it}} and Giovanni Venturi{\footnote{$^\ddagger$}
{e-mail: Armitage@bologna.infn.it}}}
\smallskip
\centerline{\it Department of Physics, University of Bologna, and}
\centerline{\it Istituto Nazionale di Fisica Nucleare,
	    Sezione di Bologna, Italy}
\bigskip
\centerline{Abstract}
{\narrower{
An extended monopole detector at constant acceleration coupled
to a massless scalar field is allowed to evolve quantum
mechanically.
It is found that while in the classical, followed by the point
particle, limit the usual result ({\sl Unruh effect}) is
recovered, in the point particle (before the classical) limit
the detector decouples from the scalar field and therefore
the effect disappears.
\smallskip}}
\par
\vskip 2truecm
\noindent
List of keywords:
\par
Accelerated observer\par
Classical limit\par
Extended monopole detector\par
Point particle limit\par
Quantum mechanics\par
Unruh effect\par
\vfill
\eject
\centerline{\bf I. Introduction}
\medskip
Hawking's remarkable discovery [1] that black holes evaporate
due to quantum particle emission and behave as if they had an
effective temperature of $(8\,\pi\, M)^{-1}$ with $M$ the mass
of the black hole also encouraged, because of the equivalence
principle, the study of field theory in accelerated systems
[2,3,4].
In particular it was found that a detector with uniform
acceleration in the usual vacuum state of flat Minkowski
space will be thermally excited to a temperature $T=a/2\,\pi$.
\par
In the derivation of the result [5] a massless, otherwise free,
scalar field ($\varphi$) is linearly coupled to a particle detector.
This then serves as a localized probe of the field which is
assumed to be in the Minkowski vacuum.
The detector is described by the DeWitt monopole moment [6] and
evolves on a trajectory of constant acceleration.
One then uses time dependent perturbation theory and
examines the two point function of the scalar field
in terms of the  detector's proper time.
A question which one can naturally ask is what happens if
one considers detectors of finite extent which evolve quantum
mechanically [7].
Such a model is considered in the next Section.
We shall use units for which $c$ and the Boltzmann constant
are set equal to unity.
\bigskip
\centerline{\bf II. An Extended Quantum Mechanical Detector}
\medskip
The Lagrangian density describing the field $\varphi$ and its
interaction with a point--like detector in four dimensional
Minkowski space will be given by:
$$
{\cal L}_\varphi=\int d\tau\,
\delta^{^{(4)}}\bigl(x^\mu-x^\mu_c(\tau)\bigr)\,
Q(\tau)\,\varphi(x^\mu)
-{1\over2}\,\eta^{\mu\nu}\,\partial_\mu\varphi\,
\partial_\nu\varphi
\ ,
\eqno(2.1)
$$
where $x^\mu_c(\tau)$ defines the world--line of the idealized
classical point--like detector, $Q(\tau)$ represents its
monopole moment at time $\tau$ and
$\delta^{^{(4)}}\bigl(x^\mu-x^\mu_c(\tau)\bigr)$ serves to transform
inertial time integrals into proper time integrals.
\par
Concerning $x^\mu_c(\tau)$ it is convenient to introduce
Rindler coordinates $\xi$ and $\tau$ (associated with the
transformation to the accelerated frame) related to the Minkowski
coordinates $x^0$ and $x^3$ through:
$$
\eqalign{
x^0 &=a^{-1}\,e^{\strut\displaystyle{a\,\xi}}\,\sinh a\tau\cr
x^3 &=a^{-1}\,e^{\strut\displaystyle{a\,\xi}}\,\cosh a\tau
\ ,\cr}
\eqno(2.2)
$$
where the acceleration is in $x^3$ direction and has a local
value equal to $a$ on the hypersurface $\xi=0$ and
$a\,e^{-a\,\xi}$ elsewhere.
In what follows we set $\xi=0$
without loss of generality, further we denote by $x^0_c$ and
$x^3_c$ the corresponding Minkowski coordinates
associated with this $\xi=0$ trajectory.
The uniformly accelerating classical observer
will then have a path given by
$x^0_c$, $x_c^3$ ($\equiv z_c$), with proper time $\tau$ and
$x^1_c(\equiv x_c)=x^2_c(\equiv y_c)=0$.
\par
In Eq. (2.1) we have the Lagrangian for the scalar field $\varphi$
and its coupling to the detector, to this we should add a
Lagrangian for the observer which leads to the desired solution
for the classical path $x^\mu_c$.
A convenient Lagrangian yielding an action whose
stationarity leads to the classical path $x_c^\mu$ is [8]
(it is sufficient to just exhibit $x^3\equiv z$
since the acceleration is in that direction,
and the corresponding Lagrangians $L_x$, $L_y$ along other
directions are obtained by setting $a=0$ and replacing $z$ by
$x$, $y$ respectively):
$$
L_z={m\over2}\,\biggl[\biggl({dz\over d\tau}\biggr)^2
+a^2\,z^2\biggr]
={m\over2}\,\biggl[{\dot z}^2
+a^2\,z^2\biggr]
\ ,
\eqno(2.3)
$$
where $m$ is the mass of the observer which is subject to a
field of force $m\,a^2\,z$ corresponding to an inverted
harmonic oscillator potential and the total Lagrangian will
of course be the sum of Eq. (2.3) with $L_x$ and $L_y$.
Further the Hamiltonian associated with $L_z$ is given by:
$$
H_z={m\over2}\,\biggl[{\dot z}^2
-a^2\,z^2\biggr]
\ ,
\eqno(2.4)
$$
which is a constant along the path.
At this point we shall further pursue the analogy with the harmonic
oscillator and if we write the total wave function for the
three-dimensional monopole detector in a factorized form
$\psi(x,\tau)\,\psi(y,\tau)\,\psi(z,\tau)$
where for each $\psi$ we shall later consider a
gaussian wavepacket, we then obtain a quantum Lagrangian density:
$$
\eqalignno{
{\cal L}_z= & |\psi(x,\tau)|^2\,|\psi(y,\tau)|^2\cr
& \times
\biggl[i\,\hbar\,\psi^\ast(z,\tau)\,\dot\psi(z,\tau)-
{\hbar^2\over2\,m}\,\biggl|{\partial\psi(z,\tau)\over\partial z}
\biggr|^2+{m\over2}\,a^2\,z^2\,|\psi(z,\tau)|^2\biggr]
\ ,
&(2.5)\cr}
$$
which in the WKB approximation leads to the Lagrangian Eq. (2.3)
[8].
We note the dynamics is associated with the $z$ coordinate
(and not $x^1\equiv x$, $x^2\equiv y$) thus again
we have just exhibited the pertinent term in the total
Lagrangian density.
The corresponding terms for $\psi(x,\tau)$ ($\psi(y,\tau)$)
are obtained from the above by setting $a=0$ and interchanging
$z$ and $x$ ($y$).
\par
The inverted harmonic oscillator has been studied [9] and in
particular the associated Green's function
$K(z,\tau;z^\prime,0)$ can be obtained from the
corresponding harmonic oscillator one through the replacement
of the frequency of oscillation by $i\,a$.
One then has:
$$
K(z,\tau;z^\prime,0)=
$$
$$
=\biggl({m\,a\over2\,\pi\,i\,\hbar\,\sinh a\,\tau}\biggr)^{1/2}\,
\exp\biggl\{{i\,m\,a\over2\,\hbar\,\sinh a\,\tau}\,
\bigl[\bigl({z^\prime}^2+z^2\bigl)\,\cosh a\,\tau-2\,z\,z^\prime
\bigr]\biggr\}
\ ,
\eqno(2.6)
$$
in which case if one begins with a gaussian wavepacket
$\psi(z^\prime,0)$:
$$
\psi(z^\prime,0)=\bigl(b\,\pi^{1/2}\bigr)^{-1/2}\,
\exp\biggl\{-\strut\displaystyle{(z^\prime-z_0)^2\over2\,b^2}
\biggr\}
\ ,
\eqno(2.7)
$$
one will obtain:
$$
\eqalignno{
\psi(z,\tau)=&{1\over\bigl(b\,\pi^{1/2}\bigr)^{1/2}}\,
\biggl({m\,a\over2\,\pi\,i\,\hbar\,\sinh a\,\tau}\biggr)^{1/2}\,
{\pi^{1/2}\over
\bigl({1\over2\,b^2}-i\,\beta\,\cosh a\,\tau\bigr)^{1/2}}\cr
&   \times
\exp\biggl\{i\,\biggl[\beta\,z^2\,\cosh a\,\tau+{\beta\over4\,b^4}
\,\biggl({1\over4\,b^4}+\beta^2\,\cosh^2 a\,\tau\biggr)^{-1}\cr
&\phantom{\times\,\exp i\ \ \ }
\times\bigl(z^2_0\,\cosh a\,\tau-2\,z\,z_0-4\,z^2\,\beta^2\,b^4\,
\cosh a\,\tau\bigr)\biggr]\biggr\}\cr
&   \times\exp\biggl\{-{\beta^2\over2\,b^2}\,
\biggl({1\over4\,b^4}+\beta^2\,\cosh^2 a\,\tau\biggr)^{-1}\,
\bigl(z-z_0\,\cosh a\,\tau\bigr)^2\biggr\}
\ ,
&(2.8)\cr}
$$
where $\beta\equiv {m\,a/(2\,\hbar\,\sinh a\,\tau)}$ and:
$$
\eqalignno{
|\psi(z,\tau)|^2 &={1\over\bigl(b\,\pi^{1/2}\bigr)}\,
{m\,a\over2\,\pi\,\hbar\,\sinh a\,\tau}\,
{\pi\over
\bigl({1\over4\,b^4}+\,\beta^2\,\cosh^2 a\,\tau\bigr)^{1/2}}\cr
&\phantom{=\ \ \ }
\times\exp\biggl\{-{\beta^2\over b^2}\,
\biggl({1\over4\,b^4}+\beta^2\,\cosh^2 a\,\tau\biggr)^{-1}\,
\bigl(z-z_0\,\cosh a\,\tau\bigr)^2\biggr\}\cr
&\equiv
{\alpha\over\sqrt\pi}\,e^{-\strut\displaystyle{\alpha^2(z-z_c)^2}}
\ ,
&(2.9)\cr}
$$
where $\alpha\equiv\beta/ \bigr[b\,
\bigl({1\over4\,b^4}+\,\beta^2\,\cosh^2 a\,\tau\bigr)^{1/2}\bigl]$
and $z_c=z_0\,\cosh a\,\tau$.
It is clear that in the limit for $\alpha\to \infty$ the R.H.S.
of Eq. (2.9) will tend to $\delta(z-z_c)$ and this suggests we
replace ${\cal L}_\varphi$ by:
$$
\eqalignno{
{\cal L}_\varphi=&\int d\tau\, \delta\bigl(x^0-x^0_c(\tau)\bigr)\,
Q(\tau)\,|\psi(x,\tau)|^2\,
|\psi(y,\tau)|^2\,|\psi(z,\tau)|^2\,\varphi(x^\mu)\cr
&-{1\over2}\,\eta^{\mu\nu}\,\partial_\mu\varphi\,
\partial_\nu\varphi
\ ,
&(2.10)\cr}
$$
where $|\psi(x,\tau)|^2$ ($|\psi(y,\tau)|^2$) is given by Eq.
(2.8) with $z_0=a=0$ and $z$ replaced by $x$ ($y$) and leads
to $\delta(x)$ ($\delta(y)$) for $\alpha(a=0)\to\infty$.
Thus in the limit for $\alpha\to\infty$,
$\alpha(a=0)\to\infty$ Eq. (2.10) leads to Eq. (2.1).
\par
As usual [5,6] we shall suppose the detector has a discrete set
of internal eigenstates described by vectors $\ket{E}$ where
$E=0$ corresponds to the ground state and evaluate the probability
$P(E)$ for the detector make a transition from the vacuum to an
excited state of energy $E$ ($\equiv\hbar\,\omega$) while the scalar
field undergoes a transition from the Minkowski vacuum to any
final state.
One then has:
$$
\eqalignno{
P(E)& =\bigl|\bra{E}\,Q(0)\,\ket{0}\bigr|^2\,
\int_0^L d\tau\,\int_0^L d\tau^\prime\,
e^{-\strut\displaystyle{i\,(\tau-\tau^\prime)\,E/\hbar}}
\, G(\tau,\tau^\prime)\cr
& =\bigl|\bra{E}\,Q(0)\,\ket{0}\bigr|^2\,\int_0^L dT\,
\int_{-L+2\,|T-L/2|}^{L-2\,|T-L/2|} dt\,
e^{-\strut\displaystyle{i\,\omega\,t}}
\, G(t,T)
\ ,
&(2.11)\cr}
$$
where, for convenience, we have considered a finite time
interval $(0,L)$ for the detector [10], defined $\tau=T+t/2$,
$\tau^\prime=T-t/2$ and:
$$
\eqalignno{
G(t,T) & ={\alpha\,\alpha^\prime\,\gamma^2\,{\gamma^\prime}^2
\over\pi^3}\,
\int dx\,dx^\prime\,dy\,dy^\prime\,dz\,dz^\prime\,
\bra{0}\,\varphi(x,y,z,x_c^0)\,
\varphi(x^\prime,y^\prime,z^\prime,{x_c^0}^\prime)\,\ket{0}
\cr
&\phantom{=\,}
\times
\exp\bigl\{-{\alpha^\prime}^2\,\bigl(z^\prime-z^\prime_c\bigr)^2
-\alpha^2\,\bigl(z-z_c\bigr)^2
-{\gamma^\prime}^2\,\bigl({x^\prime}^2+{y^\prime}^2\bigr)
-\gamma^2\,\bigl(x^2+y^2\bigr)\bigr\}\,
\cr
&={\alpha\,\alpha^\prime\,\gamma^2\,{\gamma^\prime}^2
\over4\,\pi^5}\,
\int dx\,dx^\prime\,dy\,dy^\prime\,dz\,dz^\prime\,
\exp\bigl\{
-{\gamma^\prime}^2\,\bigl({x^\prime}^2+{y^\prime}^2\bigr)
-{\gamma}^2\,\bigl(x^2+y^2\bigr)\bigr\}
\cr
&\phantom{={\alpha\,\alpha^\prime\over\pi}\int\,}
\times
{\exp\bigl\{-{\alpha^\prime}^2\,\bigl(z^\prime-z^\prime_c\bigr)^2
-\alpha^2\,\bigl(z-z_c\bigr)^2\bigr\}\over
(x-x^\prime)^2+(y-y^\prime)^2+
(z-z^\prime)^2-(x^0_c-{x^0_c}^\prime-i\,\epsilon)^2}
&(2.12)\cr}
$$
where the unprimed (primed) quantities are evaluated at $\tau$
($\tau^\prime$) and $\gamma=\alpha(a=0)$.
If we now define $u=z-z_c-z^\prime+z^\prime_c$,
$v=(z-z_c+z^\prime-z^\prime_c)/2$,
$p=x-x^\prime$, $q=(x+x^\prime)/2$,
$r=y-y^\prime$ and $s=(y+y^\prime)/2$ one obtains:
$$
\eqalignno{
G(t,T) & =-{\alpha\,\alpha^\prime\,\gamma^2\,{\gamma^\prime}^2
\over4\,\pi^5}\,
\int du\,dv\,dp\,dr\,dq\,ds\,
{\exp\bigl\{-{\alpha^\prime}^2\,(v-u/2)^2-\alpha^2\,(v+u/2)^2
\bigr\}
\over D-u^2-2\,u\,\Delta-p^2-r^2}\cr
&\phantom{=\ \ }
\times\exp\bigl\{-{\gamma^\prime}^2\,
\bigl[(q-p/2)^2+(s-r/2)^2\bigr]
-\gamma^2\bigl[(q+p/2)^2+(s+r/2)^2\bigr]\bigr\}\cr
& =-{\alpha\,\alpha^\prime\,\gamma^2\,{\gamma^\prime}^2
\over4\,\pi^5}\,
\biggl({\pi\over\alpha^2+{\alpha^\prime}^2}\biggr)^{1/2}\,
{\pi\over\gamma^2+{\gamma^\prime}^2}
\cr
&\phantom{=\ \ }
\times\int du\,dp\,dr\,
{\exp\biggl\{-\strut\displaystyle
{\alpha^2\,{\alpha^\prime}^2\over
\alpha^2+{\alpha^\prime}^2}\,u^2
-\strut\displaystyle
{\gamma^2\,{\gamma^\prime}^2\over
\gamma^2+{\gamma^\prime}^2}\,(p^2+r^2)\biggr\}
\over D-u^2-2\,u\,\Delta-p^2-r^2}
\ ,
&(2.13)\cr}
$$
where $D\equiv{4\over a^2}\,\sinh^2\bigl[{a\over2}\,(\tau-\tau^\prime)
-i\,\epsilon\bigr]$ and $\Delta\equiv z_c-z_c^\prime$.
\par
We may now consider the integral over the time variable $t$ of
$G$:
$$
\eqalignno{
&\int_{-L+2\,|T-L/2|}^{L-2\,|T-L/2|} dt\,
e^{-\strut\displaystyle{i\,\omega\,t}}\, G(t,T)=\cr
&\phantom{\int}
=-\int du\,dp\,dr\, {a^2\over4}
\int dt\, {\alpha\,\alpha^\prime\,\gamma^2\,{\gamma^\prime}^2
\over4\,\pi^5}\,
\biggl({\pi\over\alpha^2+{\alpha^\prime}^2}\biggr)^{1/2}\,
{\pi\over\gamma^2+{\gamma^\prime}^2}
\,e^{-\strut\displaystyle{i\,\omega\,t}}\cr
&\phantom{=-\int du\,}
\times{\exp\biggl\{-\strut\displaystyle
{\alpha^2\,{\alpha^\prime}^2\over
\alpha^2+{\alpha^\prime}^2}\,u^2
-\strut\displaystyle{\gamma^2\,{\gamma^\prime}^2\over
\gamma^2+{\gamma^\prime}^2}\,(p^2+r^2)\biggr\}
\over\bigl(\sinh{a\,t\over2}-u_{_+}-i\,\epsilon\bigr)\,
\bigl(\sinh{a\,t\over2}-u_{_-}-i\,\epsilon\bigr)}
\ ,
&(2.14)\cr}
$$
where we have inverted orders of integration and:
$$
\eqalignno{
u_{_\pm}& \equiv {a\over2}\,\bigl[u\,\sinh a\,T\pm
\bigl(u^2\,\sinh^2 a\,T+u^2+p^2+r^2\bigr)^{1/2}\bigr]
\cr
&=u\,{a\over2}\biggl[\sinh a\,T\pm
\biggl(\cosh^2 a\,T+{p^2+r^2\over u^2}\biggr)^{1/2}\biggr]
\ ,
&(2.15)\cr}
$$
which implies the existence of poles at (we shall close the
contour in the lower half plane):
$$
t_{_\pm}=-{2\,\pi\,i\,n\over a}+t^n_{_\pm}
\ ,
\eqno(2.16)
$$
where $n>0$ and $t^n_\pm$ is purely real and is given by:
$$
t^n_{_\pm}=(-1)^n\,{2\over a}\,\sinh^{-1}u_{_\pm}
\ .
\eqno(2.17)
$$
\par
On performing the integral with respect to $t$ in Eq. (2.14)
using the theory of residues one has:
$$
\eqalignno{
&\int_{-L+2\,|T-L/2|}^{L-2\,|T-L/2|} dt\,
e^{-\strut\displaystyle{i\,\omega\,t}}\, G(t,T)=
&(2.18)\cr
& =-{a\over2}\int\limits_{-\infty}^{+\infty}
du\,dp\,dr\,
\sum\limits_{n=1}^N\,{(-1)^n\over2\,\pi^4}\,
{e^{-\strut\displaystyle{2\,\pi\,n\over a}\,\omega}
\over u_{_+}-u_{_-}}\,
\biggl[{\delta_{_+}\,
e^{-\strut\displaystyle{i\,\omega\,t_{_+}^n}}
\over(1+u_{_+}^2)^{1/2}}
-{\delta_{_-}\,
e^{-\strut\displaystyle{i\,\omega\,t_{_-}^n}}
\over(1+u_{_-}^2)^{1/2}}\biggr]
\ ,\cr}
$$
where we have omitted an integral along the contour
which does not contribute for $L$ large and $N$ is the
largest integer $\leq(L-2\,|T-L/2|)\,a/2\,\pi$.
Further:
$$
\delta_{_\pm}=\alpha_{_\pm}\,\alpha^\prime_{_\pm}\,
\biggl({\pi\over\alpha^2_{_\pm}+
{\alpha^\prime_{_\pm}}^2}\biggr)^{1/2}\,
{\pi\over2}\,\gamma_\pm^2\,
\exp\biggl\{-{\alpha^2_{_\pm}\,{\alpha^\prime_{_\pm}}^2
\over\alpha^2_{_\pm}+{\alpha^\prime_{_\pm}}^2}\,u^2
-{\gamma_\pm^2\over2}\,(p^2+r^2)\biggr\}
\ ,
\eqno(2.19)
$$
with:
$$
\eqalignno{
\alpha_{_\pm} &= \biggl[{1\over4\,\beta_{_\pm}^2\,b^2}+
b^2\,\bigl[(-1)^n\,(1+u_{_\pm}^2)^{1/2}\,\cosh a\,T
+u_{_\pm}\,\sinh a\,T\bigr]^2\biggr]^{-1/2}
&(2.20)\cr
&\cr
\alpha_{_\pm}^\prime &=
\biggl[{1\over4\,{\beta_{_\pm}^\prime}^2\,b^2}+
b^2\,\bigl[(-1)^n\,(1+u_{_\pm}^2)^{1/2}\,\cosh a\,T
-u_{_\pm}\,\sinh a\,T\bigr]^2\biggr]^{-1/2}
&(2.21)\cr
&\cr
\beta_{_\pm} &= {m\,a\over2\,\hbar}\,
\bigl[(-1)^n\,(1+u_{_\pm}^2)^{1/2}\,\sinh a\,T
+u_{_\pm}\,\cosh a\,T\bigr]^{-1}
&(2.22)\cr
&\cr
\beta_{_\pm}^\prime &= {m\,a\over2\,\hbar}\,
\bigl[(-1)^n\,(1+u_{_\pm}^2)^{1/2}\,\sinh a\,T
-u_{_\pm}\,\cosh a\,T\bigr]^{-1}
&(2.23)\cr
&\cr
\gamma_{_\pm} &=\gamma^\prime_{_\pm}=\alpha_{_\pm}(a=0)
\ .
&(2.24)\cr}
$$
\par
Finally, for the transition probability Eq. (2.11) one will
obtain:
$$
\eqalignno{
P(E)=&\bigl|\bra{E}\,Q(0)\,\ket{0}\bigr|^2
\int_0^L dT\int\limits_0^{+\infty} du\,dp\,dr\,
{1\over\pi^4}\,
\sum\limits_{n=1}^N\,
{e^{-\strut\displaystyle{2\,\pi\,n\over a}\,\omega}
\over 2\,u\,\bigl(\cosh^2 a\,T+{p^2+r^2\over u^2}\bigr)^{1/2}}
\cr
&\times\biggl[
{\delta_{_+}\,\sin\bigl(\omega\,(-1)^n\,t_{_+}^n\bigr)
\over(1+u_{_+}^2)^{1/2}}-
{\delta_{_-}\,\sin\bigl(\omega\,(-1)^n\,t_{_-}^n\bigr)
\over(1+u_{_-}^2)^{1/2}}\biggr]
\ ,
&(2.25)\cr}
$$
and we note the appearance of the familiar Planck distribution
factor.
Concerning the above expansion we note that the first integral
may lead to a divergence as\break
$L\to\infty$ which as usual [5] can be eliminated
by considering the transition probability per unit time,
further we observe that the $u$, $p$, $r$ integrations are finite
and we may now consider the limits for $\hbar\to0$ and $b\to0$.
\par
Let us first consider $\hbar\to 0$, in such a limit one has that
$\beta_\pm$, $\beta^\prime_\pm$ tend to infinity and the Compton
wavelength of the detector will disappear from Eqs. (2.20),
(2.21) and (2.25), which will however remain a complicated
expression which nonetheless still exhibits a Planck distribution.
The expression however simplifies if we then consider the
$b\to 0$ limit (point detector), indeed in such a case:
$$
\eqalignno{
\lim_{b\to0}\,\lim_{\hbar\to0}\delta_{_\pm} &
=\lim_{b\to0} {\pi^{1/2}\over\sqrt{2}\,b\,\cosh a\,T}\,
\exp\biggl\{-\strut\displaystyle{u^2\over2\,b^2\,\cosh^2 a\,T}
\biggr\}\,{\pi\over2\,b^2}\,\exp\biggl\{-\strut\displaystyle
{\bigl(p^2+r^2\bigr)\over 2\,b^2}\biggr\}\cr
&=\pi^3\,\delta(u)\,\delta(p)\,\delta(r)
\ ,
&(2.26)\cr}
$$
where we have omitted terms of higher order in ($u,p,r$).
The transition probability $P(E)$ will then become:
$$
\lim_{b\to0}\,\lim_{\hbar\to0}P(E)=
\bigl|\bra{E}\,Q(0)\,\ket{0}\bigr|^2
\int_0^L dT\,{\omega\over2\,\pi}\,
\sum\limits_{n=1}^N\,
e^{-\strut\displaystyle{2\,\pi\,n\over a}\,\omega}
\ ,
\eqno(2.27)
$$
in agreement with previous results [5].
\par
One may now consider the limit $b\to0$, one then has:
$$
\eqalignno{
\lim_{b\to 0}\,\delta_{_\pm}=\lim_{b\to0}\,&
2\,b\,\beta_{_\pm}\,\beta^\prime_{_\pm}\,
\biggl({\pi\over\beta_{_\pm}^2+{\beta^\prime_{_\pm}}^2}
\biggr)^{1/2}\,
\exp\biggl\{-\strut\displaystyle{4\,b^2\,\beta^2_{_\pm}\,
{\beta_\pm^\prime}^2\over\beta_\pm^2+{\beta_\pm^\prime}^2}
\, u^2\biggr\}\cr
&\times 2\,\pi\,b^2\,\beta_0^2\,
e^{-\strut\displaystyle{2\,b^2\,\beta_0^2\,(p^2+r^2)}}
=0\ ,
&(2.28)\cr}
$$
where $\beta_0=m/[\hbar\,(u^2+p^2+r^2)^{1/2}]$.
Since the $u$, $p$, $r$ integrals in Eq. (2.25) are finite
and furthermore in this case so is the $T$ integral even
in the limit for $L\to\infty$, one has:
$$
\lim_{b\to0}P(E)=0
\ ,
\eqno(2.29)
$$
thus implying that a quantum--mechanically evolving point
detector at constant acceleration decouples from the scalar
field, which is rather surprising.
\bigskip
\centerline{\bf III. Conclusion}
\medskip
We have considered a massless scalar field coupled to a
monopole detector which from a given time is subjected to
a constant acceleration.
For the detector we considered a gaussian wavepacket
which we allowed to evolve according to an inverted harmonic
oscillator potential corresponding to a constant acceleration.
On examining the probability per unit time for the detector
to excite itself due to the absorption of scalar quanta,
we observed that on first considering the classical limit
($\hbar\to0$) and then the point--like limit for the
detector we reproduced the usual results.
If however one considers the point--like limit first
the detector decouples from the scalar field.
This rather surprising result is due to the fact that
once quantum--mechanical evolution is considered another
length enters the theory:
the Compton wavelength of the detector
(let us remember that we have taken $b$, $m$, $\hbar$
and $a$ as independent).
\par
Indeed in another, quite different, situation a similar
result is obtained [11].
One considers, rather than the quantum electrodynamics
of point particles, the point particle limit of a theory
of extended  particles and takes this limit on
the quantum level rather than the classical level.
Then, in the nonrelativistic approximation, such a method leads
to a vanishing self--energy and to the absence of
run--away and pre--acceleration effects.
\par
One may also ask oneself whether an approach such as the
above is possible for a black hole.
Since black hole evaporation essentially takes place
in s--waves one can make a 2-dimensional model for a black
hole and near the event horizon the Schwarzschild metric
can be put in Rindler form [2,6].
Thus one may hope that since the acceleration becomes infinite
near the event horizon it could compensate the zero due to
the point particle limit.
However for the scalar field Wightman function in two dimension
one has a logarithmic singularity rather than a pole;
usually through integration by parts one changes it to a pole,
but for our wavepacket case such an approach is considerably
more involved.
We hope to return to this point.
\vskip 1truecm
\centerline{\bf References}
\medskip
\item{[1]} S. W. Hawking, Nature {\bf 248} (1974) 30;\hfill
	   \break
	   S. W. Hawking, Comm. Math. Phys. {\bf 43} (1975)
	   199.
\par
\item{[2]} W. G. Unruh, Phys. Rev. D {\bf 14} (1976) 870.
\par
\item{[3]} P. C. W. Davies, J. Phys. A {\bf 6} (1975) 609.
\par
\item{[4]} D. W. Sciama, P. Candelas and D. Deutsch,
	   Adv. in Phys. {\bf 30} (1981) 327.
\par
\item{[5]} See for example N. D. Birrel and P. C. W. Davies,
	   {\it Quantum Fields in Curved Space} (Cambridge
	   University Press, Cambridge, England, 1982).
\par
\item{[6]} B. S. DeWitt in {\it General Relativity: an Einstein
	   Centenary Survey}, eds. by S. W. Hawking and W. Israel
	   (Cambridge University Press, Cambridge, England, 1979).
\par
\item{[7]} For another discussion of particle detectors see P. G.
	   Grove and Q. C. Ottewill, J. Phys. A {\bf 16} (1983)
	   3905.
\par
\item{[8]} R. Brout and Ph. Spindel, Nucl. Phys. B {\bf 348}
	   (1991) 405;\hfill\break
	   R. Brout, Z. Phys. B {\bf 68} (1987) 339.
\par
\item{[9]} G. Barton, Annals of Physics {\bf 166} (1986) 322.
\par
\item{[10]} In this context see {\it Response of finite time
	    particle detectors in non inertial frames and
	    curved backgrounds} by L. Sriramkumar and T.
	    Padmanabhan.
\par
\item{[11]} See H. Grotch, E. Kazes, F. Rohrlich and D.H. Sharp,
	    Acta Phys. Austr. {\bf 54} (1982) 31 and references
	    therein.
\end